\documentclass[a4paper,10pt]{article}
\usepackage{graphicx}
\usepackage{comment}
\usepackage{amssymb}
\usepackage{amsmath}
\usepackage{nicefrac}
\usepackage{graphicx}
\usepackage{dcolumn}
\usepackage{bm}
\usepackage{comment}
\usepackage{multirow}
\usepackage{cite}

\begin{document}

\huge

\begin{center}
A fast approximation to supershell partition functions
\end{center}

\vspace{0.3cm}

\large

\begin{center}
Brian Wilson\footnote{wilson9@llnl.gov}
\end{center}

\normalsize

\begin{center}
\it Lawrence Livermore National Laboratory, P.O. Box 808, L-414, Livermore, California 94551, USA
\end{center}

\vspace{0.3cm}

\large

\begin{center}
Jean-Christophe Pain$^{a,b,}$\footnote{jean-christophe.pain@cea.fr}
\end{center}

\normalsize

\begin{center}
\it $^a$CEA, DAM, DIF, F-91297 Arpajon, France\\
\it $^b$Universit\'e Paris-Saclay, CEA, Laboratoire Mati\`ere en Conditions Extr\^emes,\\
\it 91680 Bruy\`eres-le-Ch\^atel, France
\end{center}

\vspace{0.3cm}

\begin{abstract}
A formula for supershell partition functions, which play a major role in the Super Transition Array approach to radiative-opacity calculations, is derived as a functional of the distribution of energies within the supershell. It consists in an alternative expansion for an arbitrary number of electrons or holes which also allows for quick approximate evaluations with truncated number of terms in the expansion.
\end{abstract}

\section{Introduction}

The super-transition-array (STA) technique is a powerful tool to compute the radiative opacity and emissivity of intermediate to high-$Z$ plasmas in cases where the numbers of levels and lines are so important that fine-structure and detailed-configuration-accounting methods are prohibitive. The method was implemented in several opacity codes (see for instance \cite{BARSHALOM1989,BLENSKI1995,OVECHKIN2014,WILSON2015,KRIEF2021}) and successfully applied to the interpretation of absorption spectroscopy experiments, either for local-thermodynamic equilibrium (LTE) \cite{WINHART1996,BAILEY2003,BLENSKI2011,KURZWEIL2022} of non-LTE plasmas \cite{HANSEN2007,LEE2020}. It was also found to be of great interest for astrophysical applications \cite{KRIEF2016}. The STA method was significantly improved in order to refine the spectra, using an integral representation leading to the so-called Configurationally-Resolved Super Transition Array model \cite{HAZAK2012,KURZWEIL2013,KURZWEIL2016}. The utility of superconfiguration formalisms lies in the reduction of the number of atomic configurations by forming integer occupied supershells representing the entirety of configurations distributed thermally within the supershell. In most applications it is primarily applied as large supershells of Rydberg orbitals with small occupation, large total degeneracies, and orbital energy spreads about or less than the temperature of the plasma. The fundamental quantity to compute is the supershell partition function \cite{OREG1997}:
\begin{equation}
{U_Q}\left[ {\vec g} \right] \equiv \mathop {\sum\limits_{{n_1} = 0}^{{g_1}} {\sum\limits_{{n_2} = 0}^{{g_2}} {...} } }\limits_{{n_1} + {n_2} + ...{n_m} = Q} \sum\limits_{{n_m} = 0}^{{g_m}} {\left( {\begin{array}{*{20}{c}}
{{g_1}}\\
{{n_1}}
\end{array}} \right)\left( {\begin{array}{*{20}{c}}
{{g_2}}\\
{{n_2}}
\end{array}} \right)...\left( {\begin{array}{*{20}{c}}
{{g_m}}\\
{{n_m}}
\end{array}} \right)X_1^{{n_1}}X_2^{{n_2}}...X_m^{{n_m}}}
\end{equation}
where ${Q}$ is the supershell occupation, the set $\vec g = \left\{ {{g_1},{g_2},...,{g_m}} \right\}$ denotes the degeneracies of the orbitals with energies  ${\varepsilon _i}$ and
\begin{equation}
{X_i} = {e^{ - \beta \left( {{\varepsilon _i} - \mu } \right)}}.
\end{equation}
All relevant quantities, such as average occupations, transition array strengths, centroids and widths, can be calculated from them. It is worth mentioning that the main approximation of the STA approach consists in averaging the electron-electron interactions in the Boltzmann-factor, in order to deal with `` independent-electron like'' quantities. Such a treatment can be improved using the Gibbs-Bogolyubov (or Jensen-Feynman) variational approach, which enables one to include the effect of electron-electron interactions in an average manner in the one-electron energies \cite{FAUSSURIER2002,PAIN2009}.

Bar-Shalom \emph{et al.} proposed efficient recursion relations for generating these partition functions \cite{BARSHALOM1989}. The efficiency of the method was improved handling ratios of consecutive partition functions \cite{WILSON1999}. Unfortunately, in the case of high-degeneracy supershells and/or at low temperature, such relations suffer from numerical instability due to precision cancellations arising from sums of large terms of alternating sign. The instability occurs in particular when the thermal energy becomes significantly smaller than the energy spread in a supershell (\emph{i.e.} the dispersion of the energies of the subshells). In 2004, a stable and robust algorithm was proposed (see Eq. (25) in Sec. III of Ref. \cite{GILLERON2004}), relying on the computation of partition functions by nested recursion, over both the subshell and the electron numbers. The idea was to build up supershells one subshell by one subshell, at each stage from ``parent'' supershells (of one less subshell) with smaller numbers of electrons. All the terms entering the sums of this recursion are positive definite, avoiding that way cancellation issues. Three years later \cite{WILSON2007}, we proposed two main improvements of the initial method. The first one consists in extending the recursion relation to holes (complementaries of the electrons), and the second relies on the precomputation and storage of some partition functions. We recently proposed an optimization of the latter method \cite{PAIN2020}, based on the evaluation of elementary symmetric polynomials. 

We present an alternative expansion for computing supershell partition functions for an arbitrary number of electrons or holes which also allows for quick approximate evaluations with truncated number of terms in the series. In section \ref{sec2}, the expansion of the partition function is derived. It involves peculiar coefficients depending on the ``energy'' moments (the moments concern actually the Boltzmann factors), for which a recursion relation is obtained. The extension of the method to hole counting in briefly described in section \ref{sec3} and the practical implementation is explained in section \ref{sec4}. An example is finally discussed in section \ref{sec5}.

\section{Expansion of supershell partition functions in energy moments}\label{sec2}

By introducing an average or reference supershell energy, and thus an average or reference Boltzmann factor ${X_0}$, and introducing the deviation quantities
\begin{equation}
{\Delta _i} = \left( {\frac{{{X_i}}}{{{X_0}}}} \right) - 1,
\end{equation}
one can expand the definition of the partition function in powers of ${\Delta _i} $ and employ Vandermonde's identity
\begin{equation}
\left( {\begin{array}{*{20}{c}}
G\\
Q
\end{array}} \right) = \mathop {\sum\limits_{{n_1} = 0}^{{g_1}} {\sum\limits_{{n_2} = 0}^{{g_2}} {...} } }\limits_{{n_1} + {n_2} + ...{n_m} = Q} \sum\limits_{{n_m} = 0}^{{g_m}} {\left( {\begin{array}{*{20}{c}}
{{g_1}}\\
{{n_1}}
\end{array}} \right)\left( {\begin{array}{*{20}{c}}
{{g_2}}\\
{{n_2}}
\end{array}} \right)...\left( {\begin{array}{*{20}{c}}
{{g_m}}\\
{{n_m}}
\end{array}} \right)} 
\end{equation}
with the total supershell degeneracy
\begin{equation}
G = \sum\limits_{i = 1}^m {{g_i}} 
\end{equation}
to obtain a terminating series expansion
\begin{equation}
{U_Q}\left[ {\vec g} \right] = X_0^Q\left\{ {\left( {\begin{array}{*{20}{c}}
G\\
Q
\end{array}} \right) + \left( {\begin{array}{*{20}{c}}
{G - 1}\\
{Q - 1}
\end{array}} \right){\Gamma _1} + \frac{1}{2}\left( {\begin{array}{*{20}{c}}
{G - 2}\\
{Q - 2}
\end{array}} \right){\Gamma _2} + \frac{1}{6}\left( {\begin{array}{*{20}{c}}
{G - 3}\\
{Q - 3}
\end{array}} \right){\Gamma _3} + ...} \right\}
\end{equation}
where
\begin{equation}
{\Gamma _1} = \sum\limits_{i = 1}^m {{g_i}{\Delta _i}} 
\end{equation}
and
\begin{equation}
{\Gamma _2} = \left( {\sum\limits_{i = 1}^m {{g_i}{\Delta _i}} } \right)^2 - \left( {\sum\limits_{i = 1}^m {{g_i}\Delta _i^2} } \right)
\end{equation}
as well as
\begin{equation}
{\Gamma _3} = \left( {\sum\limits_{i = 1}^m {{g_i}{\Delta _i}} }\right)^3 -3\left( {\sum\limits_{j = 1}^m {{g_j}{\Delta _j}} } \right)\left( {\sum\limits_{i = 1}^m {{g_i}\Delta _i^2} } \right) +2 \left( {\sum\limits_{i = 1}^m {{g_i}\Delta _i^3} } \right)
\end{equation}
are all functionals of $\vec g$ and $\vec \Delta $.

The derivation requires nothing more than basic properties of binomial coefficients and relabelling of summations. Higher order terms ${\Gamma _m}$ can be straightforwardly obtained up to the needed value $m=Q$. The general formula can be obtained recursively (see Appendix \ref{appA}) as 
\begin{equation}\label{eq10}
\left\{ {\frac{{{\Gamma _k}}}{{k!}}} \right\} = \frac{1}{k}\sum\limits_{p = 1}^k {{{\left( { - 1} \right)}^{p + 1}}\left[ {\sum\limits_{i = 1}^m {{g_i}\Delta _i^p} } \right]} \,\left\{ {\frac{{{\Gamma _{k - p}}}}{{\left( {k - p} \right)!}}} \right\}\quad \quad \mathrm{with}\quad \quad {\Gamma _0} = 1.
\end{equation}

\section{Expansion using hole counts}\label{sec3}

Because the partition function for a given number of holes ``H'' can be written as
\begin{equation}
{U_{G - H}}\left[ {\vec g} \right] \equiv {U_G}\left[ {\vec g} \right]\left( {\mathop {\sum\limits_{{n_1} = 0}^{{g_1}} {\sum\limits_{{n_2} = 0}^{{g_2}} {...} } }\limits_{{n_1} + {n_2} + ...{n_m} = H} \sum\limits_{{n_m} = 0}^{{g_m}} {\left( {\begin{array}{*{20}{c}}
{{g_1}}\\
{{n_1}}
\end{array}} \right)\left( {\begin{array}{*{20}{c}}
{{g_2}}\\
{{n_2}}
\end{array}} \right)...\left( {\begin{array}{*{20}{c}}
{{g_m}}\\
{{n_m}}
\end{array}} \right)X_1^{ - {n_1}}X_2^{ - {n_2}}...X_m^{ - {n_m}}} } \right)
\end{equation}
where the partition function for a fully occupied supershell is given as
\begin{equation}
{U_G}\left[ {\vec g} \right] = \prod\limits_i {X_i^{{g_i}}},
\end{equation}
we note that the expression within the parenthesis can be formally expressed in moments exactly as before where now
\begin{equation}
{\Delta _i} = \left( {\frac{{{X_0}}}{{{X_i}}}} \right) - 1.
\end{equation}

\section{Practical evaluation of the partition function}\label{sec4} 

It is well known that if one expresses the exact partition function as
\begin{equation}
{U_Q}\left[ {\vec g} \right] = \left( {\begin{array}{*{20}{c}}
G\\
Q
\end{array}} \right){\left\{ {X_Q^{\mathrm{eff}}} \right\}^Q}
\end{equation}
then $ {X_Q^{\mathrm{eff}}}$ varies between
\begin{equation}
X_1^{\mathrm{eff}} \equiv \frac{1}{G}\sum\limits_{i = 1}^m {{g_i}{X_i}}
\end{equation}
and
\begin{equation}
X_G^{\mathrm{eff}} \equiv \frac{1}{G}\sum\limits_{i = 1}^m {{g_i}{\varepsilon _i}}.
\end{equation}
Choosing $ {X_0}$ within these limits does not seem to affect the numerical convergence of the expansion, but we note that if one chooses
\begin{equation}
{X_0}=X_1^{\mathrm{eff}} 
\end{equation}
then ${\Gamma _1}$ vanishes and ${\Gamma _2}$, \emph{etc.} simplifies. This simple choice enables one to obtain values of ${U_Q}$ accurately, even for large supershells with occupations up to half filling in either electron or holes.

\section{An illustrative example}\label{sec5}

\begin{figure}
\centering
\includegraphics[scale=0.7]{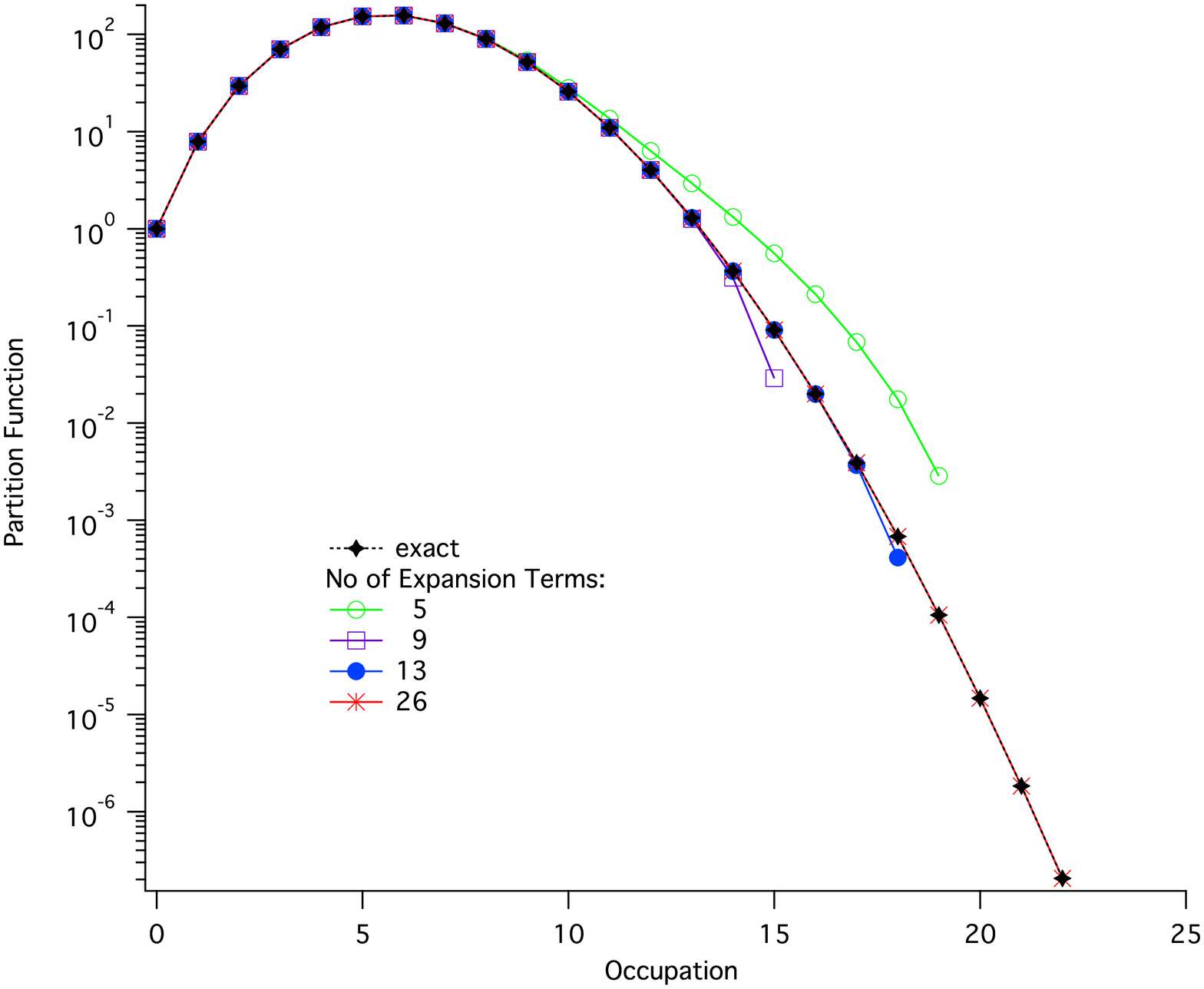}
\vspace{-10pt}
\caption{
Partition function values as a function of occupation, up to half filling of the supershell defined in table \ref{tab:table1}. Computed values where the number of expansion terms exceeds the occupation are formally exact. Using truncated expansions generates deviations from exact results, in this plot resultant negative values are omitted.}\label{fig1}
\vspace{-10pt}
\end{figure}

In this section we present the results of computing partition functions for a supershell consisting of (non-relativistic) M and N shell orbitals (see table \ref{tab:table1}).  Exact results for all occupations of the supershell are given in table \ref{tab:table2}. Values computed for 
${{{\Gamma _k}} \mathord{\left/{\vphantom {{{\Gamma _k}} {k!}}} \right.\kern-\nulldelimiterspace} {k!}}$ by recursion assuming ${X_0} =X_1^{\mathrm{eff}}$ are given in table \ref{tab:table3}.  It should be reiterated that the energy-moment expansion method presented here is formally exact if terms through the order of the occupation are retained. This was verified numerically (see Figure \ref{fig1}). The latter figure also presents the degradation of results for all occupations when the expansion is truncated. 

\begin{table}
\begin{center}
\begin{tabular}{ccc}\hline
Orbital & $\varepsilon _{n\ell}$ (eV) & ${g_{n\ell}}$\\\hline\hline
3s & -369.82378 &  2 \\
3p & -326.10399 &  6 \\
3d & -260.22501 & 10 \\
4s & -117.83349 &  2 \\
4p & -101.62248 &  6 \\
4d & -77.903611 & 10 \\
4f & -59.280040 & 14 \\\hline
\end{tabular}
\caption{\label{tab:table1}%
Supershell orbitals, energies (eV) and degeneracies for a non-relativistic treatment of a plasma copper ion at a temperature of 100.0 eV and density of 1.0 g/cm$^3$ corresponding to a chemical potential of -402.85531 eV.}
\end{center}
\end{table}

\begin{table}
\begin{center}
\begin{tabular}{cc|cc|cc}\hline
$Q$ & ${U_Q}$ & $Q$ & ${U_Q}$ & $Q$ & ${U_Q}$\\\hline\hline
  0 & 1.0000000            & 17 & 3.9041647~10$^{-3}$  & 34 & 2.9726381~10$^{-22}$ \\
  1 & 7.8738139            & 18 & 6.7916074~10$^{-4}$  & 35 & 8.9040631~10$^{-24}$ \\
  2 & 2.9521746~10$^{1}$   & 19 & 1.0552480~10$^{-4}$  & 36 & 2.3949057~10$^{-25}$ \\
  3 & 7.0225666~10$^{1}$   & 20 & 1.4680643~10$^{-5}$  & 37 & 5.7664282~10$^{-27}$ \\
  4 & 1.1914643~10$^{2}$   & 21 & 1.8327213~10$^{-6}$  & 38 & 1.2381982~10$^{-28}$ \\
  5 & 1.5367579~10$^{2}$   & 22 & 2.0570771~10$^{-7}$  & 39 & 2.3599342~10$^{-30}$ \\
  6 & 1.5684715~10$^{2}$   & 23 & 2.0794305~10$^{-8}$  & 40 & 3.9693683~10$^{-32}$ \\
  7 & 1.3019963~10$^{2}$   & 24 & 1.8958859~10$^{-9}$  & 41 & 5.8497426~10$^{-34}$ \\
  8 & 8.9669567~10$^{1}$   & 25 & 1.5609387~10$^{-10}$ & 42 & 7.4858871~10$^{-36}$ \\
  9 & 5.2013393~10$^{1}$   & 26 & 1.1616988~10$^{-11}$ & 43 & 8.2240437~10$^{-38}$ \\
 10 & 2.5710992~10$^{1}$   & 27 & 7.8209036~10$^{-13}$ & 44 & 7.6426148~10$^{-40}$ \\
 11 & 1.0932970~10$^{1}$   & 28 & 4.7652639~10$^{-14}$ & 45 & 5.8905576~10$^{-42}$ \\
 12 & 4.0300542            & 29 & 2.6283199~10$^{-15}$ & 46 & 3.6641545~10$^{-44}$ \\
 13 & 1.2960321            & 30 & 1.3122054~10$^{-16}$ & 47 & 1.7673689~10$^{-46}$ \\
 14 & 3.6559718~10$^{-1}$  & 31 & 5.9278383~10$^{-18}$ & 48 & 6.2020219~10$^{-49}$ \\
 15 & 9.0883960~10$^{-2}$  & 32 & 2.4213433~10$^{-19}$ & 49 & 1.4085403~10$^{-51}$ \\
 16 & 1.9990236~10$^{-2}$  & 33 & 8.9334309~10$^{-21}$ & 50 & 1.5538587~10$^{-54}$ \\\hline
\end{tabular}
\end{center}
\caption{\label{tab:table2}%
Exact partition function values for the supershell of table \ref{tab:table1}, as computed by the Gilleron-Pain method \cite{GILLERON2004} using quadruple precision. Values up to half filling were computed by electron recursion, those greater than half filled by hole recursion.}
\end{table}

\begin{table}
\begin{center}
\begin{tabular}{cc|cc}\hline
$k$ & ${{{\Gamma _k}} \mathord{\left/{\vphantom {{{\Gamma _k}} {k!}}} \right.\kern-\nulldelimiterspace} {k!}}$ & $k$ & ${{{\Gamma _k}} \mathord{\left/{\vphantom {{{\Gamma _k}} {k!}}} \right.\kern-\nulldelimiterspace} {k!}}$\\\hline\hline
  1 &   0.0000000          & 14 &   1.4915325~10$^{6}$ \\
  2 &  -3.4548373~10$^{1}$ & 15 &  -1.8116095~10$^{6}$ \\
  3 &   4.0837025~10$^{1}$ & 16 &  -1.1444408~10$^{6}$ \\
  4 &   4.9174449~10$^{2}$ & 17 &   4.6666847~10$^{6}$ \\
  5 &  -1.1487957~10$^{3}$ & 18 &  -3.0238046~10$^{6}$ \\
  6 &  -3.1481839~10$^{3}$ & 19 &  -3.7834926~10$^{6}$ \\
  7 &   1.3204391~10$^{4}$ & 20 &   7.6013640~10$^{6}$ \\
  8 &   2.9342658~10$^{3}$ & 21 &  -2.7294851~10$^{6}$ \\
  9 &  -7.6720735~10$^{4}$ & 22 &  -5.2354154~10$^{6}$ \\
 10 &   8.5099748~10$^{4}$ & 23  &  6.6968311~10$^{6}$ \\
 11 &   2.0615245~10$^{5}$ & 24 &  -9.8015892~10$^{5}$ \\
 12 &  -5.5908703~10$^{5}$ & 25 &  -4.0814635~10$^{6}$ \\
 13 &   3.5546876~10$^{4}$ & 26 &   3.5296876~10$^{6}$ \\\hline
\end{tabular}
\end{center}
\caption{\label{tab:table3}%
Values of $ {\Gamma _k}$ divided by ${k!}$ calculated by recursion assuming ${X_0} =X_1^{\mathrm{eff}}$ for occupation up to half filling of the supershell defined in table \ref{tab:table1}. Values can be negative as well as positive, and depend on the choice of ${X_0}$.}
\end{table}

\section{Summary} 

We have presented an alternative expansion for computing supershell partition functions for an arbitrary number of electrons or holes. It involves binomial coefficients and specific coefficients which can be determined by a recursion relation involving moments of the distribution of energies (more precisely of Boltzmann factors) within the supershell. Truncating the number of terms in the expansion enables one to perform a fast approximate computation of the partition functions, which are the cornerstone of the Super Transition Array method for radiative-opacity calculations.

\section*{Acknowledgements}
This work performed under the auspices of the U.S. Department of Energy by Lawrence Livermore National Laboratory under Contract DE-AC52-07NA27344.

\appendix

\section{Derivation of the recursion relation for coefficients $\Gamma_k$ (equation (\ref{eq10}))}\label{appA}

If we perform the summation not over the populations $0 \le {n_i} \le {g_i}$ but, which is equivalent and more convenient, over the G microstates (the population of a microstate being ${s_i}$= 0 or 1), then we get
\begin{equation}
{U_Q}\left[ {\vec g} \right] = \underbrace {X_0^G\sum\limits_{{s_1} = 0}^{} {\sum\limits_{{s_2} = 0}^{} { \cdots \sum\limits_{{s_G} = 0}^{} {\prod\limits_{i = 1}^G {{{\left( {1 + {\Delta _i}} \right)}^{{s_i}}}} } } } }_{{s_1} + {s_2} +  \cdots {s_G} = Q}
\end{equation}
and one can use
\begin{equation}
\prod\limits_{i = 1}^G {{{\left( {1 + {\Delta _i}} \right)}^{{s_i}}}}  = \prod\limits_{i = 1}^G {\left( {1 + {s_i}{\Delta _i}} \right)}
\end{equation}
which enables one to write, after expanding the product in the right-hand-side of the preceding equation:
\begin{equation}
{U_Q}\left[ {\vec g} \right] = X_0^G\sum\limits_{k = 0}^Q {\left( {\begin{array}{*{20}{c}}
{G - k}\\
{Q - k}
\end{array}} \right)\,} {\Phi _k},
\end{equation}
where \cite{PAIN2011}
\begin{equation}
{\Phi _k} = \frac{1}{{k!}}\,{\left. {{{\left( {\frac{\partial }{{\partial z}}} \right)}^k}\prod\limits_{i = 1}^G {\left( {1 + {\Delta _i}z} \right)} } \right|_{z = 0}},
\end{equation}
which can be put in the form
\begin{equation}
{\Phi _k} = \frac{1}{{k!}}\,{\left. {{{\left( {\frac{\partial }{{\partial z}}} \right)}^{k - 1}}\left( {\frac{\partial }{{\partial z}}} \right)\prod\limits_{i = 1}^G {\left( {1 + {\Delta _i}z} \right)} } \right|_{z = 0}}
\end{equation}
yielding
\begin{equation}
{\Phi _k} = \frac{1}{{k!}}{\left( {\frac{\partial }{{\partial z}}} \right)^{k - 1}}{\left. {\sum\limits_{i = 1}^G {{\Delta _i}} \prod\limits_{j = 1,j \ne i}^G {\left( {1 + {\Delta _j}z} \right)} } \right|_{z = 0}}
\end{equation}
or
\begin{equation}
{\Phi _k} = \frac{1}{{k!}}{\left. {{{\left( {\frac{\partial }{{\partial z}}} \right)}^{k - 1}}\sum\limits_{i = 1}^G {{\Delta _i}\frac{{\prod\limits_{j = 1}^G {\left( {1 + {\Delta _j}z} \right)} }}{{\left( {1 + {\Delta _i}z} \right)}}} } \right|_{z = 0}}.
\end{equation}
Using the Leibniz rule, we get
\begin{equation}
{\Phi _k} = \frac{1}{{k!}}\sum\limits_{i = 1}^G {{\Delta _i}\sum\limits_{p = 0}^{k - 1} {\left( {\begin{array}{*{20}{c}}
{k - 1}\\
p
\end{array}} \right)} } {\left. {\left\{ {{{\left( {\frac{\partial }{{\partial z}}} \right)}^{k - 1 - p}}\prod\limits_{j = 1}^G {\left( {1 + {\Delta _j}z} \right)} } \right\}\left\{ {{{\left( {\frac{\partial }{{\partial z}}} \right)}^p}\frac{1}{{\left( {1 + {\Delta _i}z} \right)}}} \right\}} \right|_{z = 0}};
\end{equation}
setting $p' = p + 1$ and using
\begin{equation}
{\left. {{{\left( {\frac{\partial }{{\partial z}}} \right)}^p}\frac{1}{{\left( {1 + {\Delta _i}z} \right)}}} \right|_{z = 0}} = {\left( { - 1} \right)^p}p!\Delta _i^p
\end{equation}
one gets
\begin{equation}
{\Phi _k} = \frac{1}{{k!}}\sum\limits_{p' = 1}^k {\left[ {\sum\limits_{i = 1}^G {\Delta _i^{p'}} } \right]} \left( {\begin{array}{*{20}{c}}
{k - 1}\\
{p' - 1}
\end{array}} \right){\left( { - 1} \right)^{p' - 1}}\left( {p' - 1} \right)!{\left. {\left\{ {{{\left( {\frac{\partial }{{\partial z}}} \right)}^{k - p'}}\prod\limits_{j = 1}^G {\left( {1 + {\Delta _j}z} \right)} } \right\}} \right|_{z = 0}}
\end{equation}
and finally, since
\begin{equation}
\sum\limits_{i = 1}^G {\Delta _i^p}  = \sum\limits_{i = 1}^m {{g_i}\Delta _i^p}
\end{equation}
we obtain
\begin{equation}
{\Phi _k} = \frac{1}{k}\sum\limits_{p = 1}^k {{{\left( { - 1} \right)}^{p + 1}}\left[ {\sum\limits_{i = 1}^m {{g_i}\Delta _i^p} } \right]} \,{\Phi _{k - p}}.
\end{equation}
The ${\Phi _k}$ coefficients are the ${\Gamma _k}$ divided by $k!$, which means
\begin{equation}
{\Gamma _k} = \sum\limits_{p = 1}^k {{{\left( { - 1} \right)}^{p + 1}}\left( {p - 1} \right)!\left( {\begin{array}{*{20}{c}}
{k - 1}\\
{p - 1}
\end{array}} \right)\left[ {\sum\limits_{i = 1}^m {{g_i}\Delta _i^p} } \right]} \,{\Gamma _{k - p}}
\end{equation}
or in a more convenient form for the expansion of ${U_Q}$ 
\begin{equation}
\left\{ {\frac{{{\Gamma _k}}}{{k!}}} \right\} = \frac{1}{k}\sum\limits_{p = 1}^k {{{\left( { - 1} \right)}^{p + 1}}\left[ {\sum\limits_{i = 1}^m {{g_i}\Delta _i^p} } \right]} \,\left\{ {\frac{{{\Gamma _{k - p}}}}{{\left( {k - p} \right)!}}} \right\}\quad \quad {\Gamma _0} = 1.
\end{equation}

\end{document}